# Wafer-Scale Growth of Sb₂Te₃ Films via Low-Temperature ALD

# for Self-Powered Photodetector

*Jun Yang[1,2*], Jianzhu Li[3], Amin Bahrami[1], Noushin Nasiri[4], Sebastian Lehmann[1], Magdalena Ola Cichocka[1], Samik Mukherjee[1*], Kornelius Nielsch[1,2*]*

[1] *Institute for Metallic Materials, Leibniz Institute of Solid State and Materials Science, 01069 Dresden, Germany*

[2] *Institute of Materials Science, Technische Universität Dresden, 01062 Dresden, Germany*

[3] *School of Materials Science and Engineering, Harbin Institute of Technology (Weihai), Weihai, West Road 2, Weihai, Shandong 264209, China*

[4] *School of Engineering, Faculty of Science and Engineering, Macquarie University, Sydney, NSW 2109, Australia*

## Abstract

In this work, we demonstrate the performance of a silicon-compatible high-performance self-powered photodetector. A wide detection range from visible (405 nm) to near-infrared (1550 nm) light was enabled by the vertical p-n heterojunction between the p-type antimony telluride (Sb₂Te₃) thin film and the n-type silicon (Si) substrates. A Sb₂Te₃ film with a good crystal quality, low density of extended defects, proper stoichiometry, p-type nature, and excellent uniformity across a 4-inch wafer was achieved by atomic layer deposition at 80 °C using (Et₃Si)₂Te and SbCl₃ as precursors. The processed photodetectors have a low dark current (~20 pA), a high responsivity of (~4.3 A/W at 405 nm and ~150 mA/W at 1550 nm), a peak detectivity of ~$1.65 \times 10^{14}$ Jones, and a quick rise time of ~98 μs under zero bias voltage. Density functional theory calculations reveal a narrow, near-direct, type-II bandgap at the heterointerface that supports a strong built-in electric field leading to efficient separation of the photogenerated carriers. The devices have long-term air stability and efficient switching behaviour even at elevated temperatures. These high-performance self-powered p-Sb₂Te₃/n-Si heterojunction photodetectors have immense potential to become reliable technological building blocks for a plethora of innovative applications in next-generation optoelectronics, silicon-photonics, chip-level sensing and detection.

---

[*] Corresponding authors: Kornelius Nielsch (k.nielsch@ifw-dresden.de); Samik Mukherjee (s.mukherjee@ifw-dresden.de); Jun Yang (j.yang@ifw-dresden.de)

# 1 Introduction

Photodetectors are semiconducting optoelectronic devices that can convert light signals into electric signals. A broadband detection range of photodetectors is essential to meet various requirements like environmental monitoring, sensing applications, optical communication, and medical diagnoses.[1] In addition, compared to a conventional photodetector device, which can operate only under an external bias, a self-powered photodetector can operate without any power supply. Consequently, the latter is more suited to work in conditions that impose stringent requirements of ultralow power consumption, small device footprint, and wireless condition.[2-4] Since the successful separation of mono-layer graphene in 2004, two-dimensional (2D) materials have been extensively developed due to their good light absorption, high charge carrier mobility, and excellent photoresponse properties.[5] The 2D materials feature a distinctive 2D planar atomic structure that makes them more stable than other semiconductor materials and allows for a large effective surface area. Additionally, as the surface of 2D materials is naturally passivated without any dangling bonds, van der Waals interactions enable the construction of vertical heterostructures using different 2D or 3D materials without "lattice mismatch" issues.[6] Recently, topological insulator (TI)-based 2D materials have attracted considerable attention owing to their tuneable optical properties, ultrafast carrier dynamics, and low power dissipation property.[7] As an additional paradigm, the metallic Dirac surface states, protected by time-reversal symmetry, offer topological protection to carrier backscattering, resulting in ultrafast carrier transport and long-term chemical stability in air.[8]

To date, photodetectors based on 2D layered TI materials have primarily focused on nanoflakes or nanoplate structures. Seldom do mechanical exfoliation or solvothermal strategies to synthesize materials lead to conformal growth that covers the entire wafer.[9, 10] This limits their scalability for applications that demand large-size focal plane array devices. Additionally, a majority of these aforementioned studies report peak responsivity values that are in the ~mA/W range. The responsivity of a photodetector, which reflects the photodetector gain, is indicative of the material quality of the active region. A material with an unsatisfactory crystalline quality causes a reduction in the minority carrier recombination lifetime, leading to poor responsivity values. Crystal defects in the active region also contribute to the overall larger dark current that adds to the noise equivalent power of a photodetector, diminishing its detectivity. On the other hand, owing to the excellent quality of the films, TI-based photodetectors grown using chemical vapour deposition (CVD) or molecular beam epitaxy (MBE) techniques often show far superior performance with responsivity values reaching the ~A/W range.[11-15] Nonetheless, CVD- and MBE-grown TI films also suffer from several issues, such as limited wafer-scale coverage,[12-14] stringent substrate requirements that are not scalable,[11, 13] or high thermal budgets that can prevent large-scale integration using complementary metal-oxide-semiconductor (CMOS) processing.[16] In this study, the atomic layer deposition (ALD) technique was used to grow p-type $Sb_2Te_3$ TI films at a low temperature of 80 °C on an n-type Si substrate, thereby forming a vertical p-n heterojunction. The vertical design would reduce the diffusion length and depletion dimensions for photogenerated carriers, resulting in a quicker photoresponse time. A schematic illustration is shown in Figure S1. Owing to its sequential and self-limiting surface reactions, ALD provides a high conformality and atomic-scale precision in controlling the layer thicknesses.[17, 18] Due to the excellent crystal quality of the as-grown TI films, the photodetector has excellent self-powered performance with a broadband photodetection range (405 − 1550 nm), a remarkable peak responsivity (∼4.3 A/W at 405 nm, under a low power density of 7.38 μW/cm²), a low dark current (in the ∼pA regime), and high detectivity (in the ∼$10^{14}$ Jones regime), without the need for any postdeposition thermal annealing steps. In addition, the photodetectors demonstrate good stability and repeatability, even at temperatures as high as 473 K, demonstrating the device's viability for applications in logic circuits used in harsh conditions. The wafer-scale array was also fabricated on a 4-inch n-type Si substrate and showed high uniformity and scalability. The findings reported in this work indicate that ALD-processed TI/Si heterostructures could offer an effective alternative for the fabrication of self-powered, cost-effective, scalable, and CMOS-compatible broadband photodetectors.

## 2   Experimental Section

### a.   ALD growth of Sb₂Te₃ thin films

The $Sb_2Te_3$ thin film was deposited using a thermal ALD reactor (Veeco Savannah S200) at a low deposition temperature of 80 °C. $(Et_3Si)_2Te$ and $SbCl_3$ were used as precursors. High-purity $N_2$ was used as the carrier gas, and the flow in the chamber was maintained at a rate of 10 sccm during the ALD reaction process. The optimized pulse and purge times for one ALD deposition cycle $((Et_3Si)_2Te/N_2/SbCl_3/N_2)$ were 0.5/10/0.5/10 s. The Si substrates were cleaned with acetone, ethanol, and deionized water in an ultrasonic bath for 15 min before use. Finally, the samples were dried using a nitrogen gun before loading into the ALD chamber.

### b.   Device fabrication

All photolithography steps were performed using a laser writer (μPG 101, Heidelberg Instruments GmbH, Germany, 375 nm irradiation wavelength). For the first step, a photoresist (AZ10XT, MicroChemicals GmbH, Germany) and developer (AZ400K, MicroChemicals GmbH, Germany) were used to place pattern alignment markers and deposit the $Al_2O_3$ insulator layer on the n-type Si substrate. Subsequently, the photoresist was removed using n-methyl pyrrolidone (NMP). Then, a 30 nm $Sb_2Te_3$ thin film was deposited at 80 °C via ALD using the same photolithographic process. Finally, the photoresist (AZ5214E, MicroChemicals GmbH, Germany) and developer (AZ 726 MIF, MicroChemicals GmbH, Germany) were used to structure the top contact on the $Sb_2Te_3$ layer, and the Ti (10 nm)/Au (90 nm) electrode was grown by sputtering. The photoresist was removed by NMP. The opening window is 1×1 mm. The detailed process is shown in Figure S1.

### c. Characterization of $Sb_2Te_3$ thin film and photodetector

The crystallinity and composition of $Sb_2Te_3$ thin film were characterized using X-ray diffraction (XRD, with Co K$\alpha$ radiation, D8 advance, Germany), X-ray reflectometry (XRR, X'Pert MRD PRO), Raman spectroscopy (InVia Raman microscopes with the excitation wavelength at 442 nm), field emission scanning electron microscopy (FE-SEM, Sigma300-ZEISS FESEM), X-ray photoelectron spectroscopy (XPS, Thermo Scientific K-Alpha+) and TEM (double-corrected FEI Titan 80-300 microscope operated at an acceleration voltage of 300 kV; the lamella was prepared by FIB). The electrical resistivity, conductivity, and carrier concentration values of the thin film were measured with a Linseis Thin Film Analyser. The electrical performance of the photodetector device was measured using a Keithley 2450 source metre and probe station under different laser wavelengths.

### d. Theoretical Calculations

All the calculations were made using the Vienna ab initio Simulation Package (VASP) code, which is based on density functional theory (DFT). For the exchange-correlation function, the generalized gradient approximation (GGA) with Perdew–Burke–Ernzerhof (PBE) was applied for structural relaxation. The cut-off energy for the plane wave basis was 300 eV. The $1 \times 5 \times 1$ k-point grid generated by the Gamma k scheme was used for structural optimization and energy calculations. The vacuum thickness was fixed to 15 Å. The lattice mismatch was less than 5% in this supercell, which contained 56 Si atoms, 24 Sb atoms, and 36 Te atoms. The force and energy convergence criteria were set as 0.01 eV Å$^{-1}$ and $10^{-4}$ eV, respectively.

## 3   Results and Discussion

The growth mechanism and optical images of $Sb_2Te_3$ thin film on a 4-inch wafer via ALD are shown in Figures S2 and S3 (Supporting Information). $(Et_3Si)_2Te$ and $SbCl_3$ were used for the ALD reaction at a low deposition temperature of 80 °C, achieving steady-state growth per cycle of 0.19 Å. $Sb_2Te_3$ films are known to possess a rhombohedral phase with a space group of *R-3m*. It is formed by stacking five-atom layers, which is known as a quintuple layer (QL).[19, 20] Each individual QL has atoms covalently bonded *via* van der Waals forces in the following atomic arrangement: $Te^{(1)}$-Sb-$Te^{(2)}$-Sb-$Te^{(1)}$, as shown schematically in Figure 1a. This out-of-plane van der Waals contact between layers without surface dangling bonds can reduce carrier generation-recombination noise. The X-ray diffraction (XRD) peaks shown in Figure 1b demonstrate that the $Sb_2Te_3$ thin film has a preferred orientation along the *c*-axis direction. To gather further insights into the crystal quality of the layers, high-resolution TEM (HRTEM) measurements were carried out. Figure 1c displays the cross-sectional HRTEM image of the $Sb_2Te_3$ layer, recorded along the [-2-10] crystallographic direction. The data confirm the good crystal quality of the as-grown film and the lack of any extended defects. The top-view scanning electron microscope (SEM) image with energy dispersive X-ray (EDX) mapping shown in Figure 1d indicates that the $Sb_2Te_3$ thin film has a continuous structure, which is crucial for good electrical conductivity. Three typical peaks, characteristic of crystalline $Sb_2Te_3$, were observed in the Raman spectrum (Figure 1e). The peaks centred at ~69.6 and ~169.9 $cm^{-1}$ could be assigned to the $A_{1g}(1)$ and $A_{1g}(2)$ phonon modes, respectively. $A_{1g}(1)$ mostly involves the symmetrical out-of-plane vibrations of the Sb-Te atoms in opposite directions, while the $A_{1g}(2)$ mode is associated with the relative vibrations between the Te and Sb atoms in the same direction. The peak centred at ~112.3 $cm^{-1}$ is ascribed to the in-plane $E_g(2)$ mode.[21] The assigned peak positions are in good agreement with the theoretical Raman spectra of $Sb_2Te_3$ thin films.[22] X-ray photoelectron spectroscopy (XPS) characterization was also performed on the $Sb_2Te_3$ thin film (Figures 1f and 1 g). The Sb $3d_{3/2}$ and Sb $3d_{5/2}$ peaks can be identified at ~539.6 and ~530.4 eV, respectively. The peaks centred at ~582.8 and ~572.4 eV are attributed to Te $3d_{3/2}$ and Te $3d_{5/2}$, respectively. In addition, the atomic ratio of Sb/Te is close to 2:3, confirming the stoichiometric composition of $Sb_2Te_3$. The Hall resistance versus magnetic field of the $Sb_2Te_3$ thin film is shown in Figure 1h. The p-type nature of the $Sb_2Te_3$ thin film was demonstrated by its positive Hall coefficient of 134 $cm^3$ $C^{-1}$. The p-type nature likely originates from the narrow bandgap, leading to thermal generation of free carriers and the presence of vacancies within the film.

To explore the light detection properties of the $Sb_2Te_3$/Si heterostructure, photodetector devices were fabricated (see the experimental sections). Figure 2a shows the current-voltage (*I-V*) characteristics of the fabricated device as a function of the illumination light density (intensity) at 405 nm wavelength (λ). The photocurrent strongly depends on the laser power density. A clear rectifying behaviour was observed, demonstrating the existence of a sharp p-n junction at the interface between $Sb_2Te_3$ and Si.[13] An extremely low dark current of 20 pA was measured under zero bias (Figure 2b), which is three orders of magnitude less than previously reported for photodetectors based on topological insulator materials.[23, 24] This ultralow dark current is attributed to the low density of extended defects within the TI and a clean interface between the $Sb_2Te_3$ and the Si,[7, 25] leading to a low bulk dark current, as well as to the lack of dangling bonds at the surface of the 2D-TI, resulting in a low surface dark current without the need for an additional surface passivating layer. The maximum electrical output power ($P_{el,max}$) of 3.36 μW was generated at 0.09 V (Figure 2c) with an incident power density of 26.81 mW/cm$^2$, indicating that the most suitable output working voltage of this photodetector device is 0.09 V. Figure 2d shows that the open circuit voltage ($V_{oc}$) as a function of power density saturates at ~1.65 V under an illumination wavelength of 405 nm at room temperature. The measured $V_{oc}$ starts from a small value at a low illumination density due to the small density of photogenerated charge carriers. As the concentration of the photogenerated charge carriers increases with increasing illumination power density, $V_{oc}$ also shows a sharp rise to 1.57 mV at 21.81 mW/cm$^2$. However, a further increase in the power density results in the saturation of $V_{oc}$ at ~1.65 V, possibly due to the domination of the Auger recombination process.[26] A similar saturation behaviour of the $V_{oc}$ was previously reported for $WSe_2$ p-i-n photodetectors.[27] The inset in Figure 2(d) shows the photocurrent under different incident laser power densities. The linear relationship between the photocurrent ($I_{ph}$) and incident optical power density ($P$) in a double-log coordinate can be fitted by the power law, $I_{ph} \propto P^\alpha$,[28] From the fit, α was determined to be 0.81 at λ = 405 nm at 0 bias voltage. A value of α less than unity usually indicates the presence of trap states between the Fermi level and the band edges.[29]

Figures 3a-c and Figure S4 analyse the switching behaviours of the $Sb_2Te_3$/Si photodetector ranging from 405 to 1550 nm at zero bias voltage. The photodetector could ideally be switched between its "ON" and "OFF" states by periodically turning the laser on and off. Under a pulsed illumination power density of 26.81 mW/cm$^2$ at 405 nm, a high photocurrent of 61.6 μA was achieved. A similar photocurrent was also obtained at 532 nm under a power density of 23.39 mW/cm$^2$ (Figure S4). However, with a further increase in the detection wavelength, the photocurrent showed a consistent decrease. For instance, photocurrent values of 28.92 μA at 850 nm (26.41 mW/cm$^2$), 11.12 μA at 980 nm (26.81 mW/cm$^2$), and 6.37 μA at 1550 nm (36.65 mW/cm$^2$) were measured. The reason is that the absorption intensity of $Sb_2Te_3$ weakens with increasing light wavelength.[13] The photoresponse speed of the photodetector is shown in Figure 3d. The rise ($t_{rise}$) and fall times ($t_{fall}$) were computed from the time taken for the photocurrent to increase from 10% to 90% of the peak value, and vice versa. The extracted values of $t_{rise}$ and $t_{fall}$ were 98 and 133 μs, respectively. This fast dynamic response can be attributed to the efficient separation of photogenerated charge carriers by a strong built-in electric field at the heterointerface. The rise time is slightly shorter than the fall time. This behaviour could be attributed to the difference in carrier mobilities and therefore the different transit times of carriers across the $Sb_2Te_3$ film and Si substrate.

The responsivity (*R*) and detectivity (*D\**) of the $Sb_2Te_3$/Si heterostructure photodetector under different wavelengths were calculated according to the following equations:

$$R = \frac{I_{ph}}{P \cdot A} \qquad (1)$$

$$D^* = \frac{\sqrt{A}}{NEP} \qquad (2)$$

where $I_{ph}$ is the photocurrent, *P* is the incident optical power density, *A* is the effective illuminated area, which is equal to the device active area since the laser spot size was larger than the device cross-section, and *NEP* is the noise equivalent power. NEP can be expressed as:

$$NEP = \frac{I_{rms}}{R\sqrt{\Delta f}} \qquad (3)$$

where $\Delta f$ is the noise equivalent bandwidth and $I_{rms}$ is defined as:

$$I_{rms} = \sqrt{\left(I_{therm}^2 + I_{shot}^2\right)} \qquad (4)$$

$I_{therm}$ and are $I_{shot}$ the thermal noise and shot noise, respectively.[30] The contributions of the thermal and shot noises to $I_{rms}$ were obtained using $I_{therm} = \sqrt{4kT\Delta f / R_{shunt}}$, where the shunt resistance ($R_{shunt}$) was determined from the first derivative of the bias-to-dark current near 0 V, and $I_{shot} = \sqrt{2qI_{dark}\Delta f}$, where $I_{dark}$ is the dark current. The calculated values of *R* and *D\** (for a noise equivalent bandwidth or $\Delta f$ of 1 Hz) values are illustrated in Figures 3e and S5, respectively. The peak *R* value of 4.287 A/W, obtained under a low incident power density of 7.38 μW/cm$^2$ ( $\lambda$ = 405 nm) under zero bias, is the highest responsivity reported thus far for self-powered photodetectors. A comparison between the peak $R$ values obtained in this work and those reported elsewhere in the literature is also depicted graphically in Figure 3f.[13, 31-50] For the different wavelengths shown in Figure 3e, $R$ can be seen to decrease progressively with increasing power density. This could be attributed to an excess photogeneration of carriers at high incident power levels, possibly leading to an increased rate of Auger recombination process and therefore a reduction in the photocurrent. In addition, a high detectivity of $1.65 \times 10^{14}$ Jones was achieved under similar conditions (Figure S5), which can be attributed to the low dark current measured under zero bias (Figure 2b). These performance parameters are much better than those of most traditional semiconductor-based photodetectors (see Table 1). Furthermore, the fabricated $Sb_2Te_3$/Si photodetector demonstrated outstanding stability and repeatability, as presented in Figures 3g-h. The measured photocurrent remained constant even after one million cycles of operation. Figure 3h analyses the switching behaviour of the photodetector after one week, four weeks, and twelve weeks. The photocurrent and switching behaviours are identical even after 12 weeks of storage, suggesting the excellent stability and robustness of the ALD-synthesized $Sb_2Te_3$/Si heterostructure-based device.

The band alignment is crucial in heterojunction structured devices since it dictates the transport of carrier heterointerfaces of the two types of materials.[51] The band structures of the $Sb_2Te_3$/Si heterostructure were calculated using density functional theory (DFT), as shown in Figure 4a. The contributions from the Si, Sb, and Te atoms are indicated by the relative sizes of cyan, brown, and green circles, respectively. The fundamental bandgap, as extracted from the DFT calculations, is 0.15 eV, which is smaller than the bandgaps of both $Sb_2Te_3$ (0.29 eV) and Si (1.12 eV).[52] The nature of the bandgap at the heterointerface is quasi-direct since the direct transition (single arrow in Figure 4a) is only ~20 meV above the indirect transition. The conduction band minimum (CBM) and valence band maximum (VBM) of the heterostructure are predominantly contributed by $Sb_2Te_3$ and Si, respectively, thereby forming a type-II heterointerface. Such a composite band structure originates not only from the overlap of the individual energy bands of $Sb_2Te_3$ films and bulk Si but also from subtle structural effects such as charge redistribution among the different atomic constituents. To elucidate the same, the differential charge density was investigated using DFT to distinguish the distribution of charge at the interface of $Sb_2Te_3$ and Si (Figure 4b and Figure S6, Supporting Information). The yellow- and purple-coloured regions designate the accumulation and depletion of electronic charges, respectively. Due to the charge reconfiguration, the electrons can be seen to accumulate on the Te atoms while being depleted from the topmost layer of the Si atoms and the Sb atoms within the $Sb_2Te_3$ film. Such polarization effects, brought about by the redistribution of the electronic charges, impact the resultant band structure during heterostructure formation. Such phenomenon of charge redistribution is not unique to the $Sb_2Te_3$/Si system and was also observed on other heterostructured materials.[53, 54] The energy band alignment under illumination was also further studied. This band bending results in a strong built-in electric field directed from the n-type Si to the p-type $Sb_2Te_3$, as illustrated in Figure 4c. Those carriers that are generated within the depletion region become spatially separated under the influence of the strong built-in field. The recombination probability of the electron-hole pairs is also reduced due to the presence of the type II band alignment at the heterointerface.[55] The holes drift into the p-$Sb_2Te_3$ layer, and the electrons drift into the n-Si substrate. Outside the region of the built-in electric field, the carriers diffuse under a concentration gradient until they reach the metal contact and are collected at the electrodes (Figure 4d). While the high-quality and low defect density ALD-synthesized $Sb_2Te_3$ films ensure a reduced rate of Shockley-Reed-Hall recombination, the unique band structure and the topological surface states provide an efficient conducting channel to the photogenerated holes for collection, resulting in an ultrafast photoresponse without an external bias voltage. A comprehensive discussion, based on semi-classical approach, regarding the interaction of electromagnetic waves with electrons/holes, is provided in the Supporting Information.

The temperature-dependent performance of the $Sb_2Te_3$/Si heterostructure is shown in Figure 5a. The device was heated on a hot plate (298 to 473 K) and illuminated with a 523 nm laser with a power density of 23.39 mW/cm$^2$. The photodetector shows stable ON/OFF behaviour even at a temperature as high as 473 K, which confirms its suitability for logic circuit applications in harsh environments. However, a gradual decrease in the photocurrent was observed as the operating temperature increased. The photo-to-dark current ratio (PDCR) is defined as

$$PDCR = \frac{I_{\text{ph}} - I_{\text{dark}}}{I_{\text{dark}}} \qquad (5)$$

It was observed that the PDCR values diminished as the temperature increased (Figure 5b), which could be attributed to the continuous decrease in the photocurrent and the increases in dark current. The performance degradation could be explained by the thermoelectric property of $Sb_2Te_3$, as shown in Figure 5c. The resistivity increases from 49.62 to 72.22 $\mu\Omega \cdot$m, and the carrier concentration falls from $7.2 \times 10^{18}$ to $5.3 \times 10^{18}$ cm$^{-3}$ with increasing temperature, ranging from 298 to 473 K. The low photocurrent of the device at high temperatures could have the following possible explanations. First, the phonon scattering becomes stronger as the operating temperature increases, which increases the recombination of photogenerated electrons and holes. Second, the absorption of $Sb_2Te_3$ decreases with increasing temperature in almost all wavelength ranges.[56] Beyond 473 K, the onset of the evaporation of the Te atoms from the $Sb_2Te_3$ thin film was marked, and consequently, the introduction of structural defects within the layer during this heating process further degraded the device performance. The wafer-scale array was also built on a 4-inch n-type Si substrate to explore the uniformity and scalability of $Sb_2Te_3$/Si photodetector devices, as shown in Figure 5d. Histograms showing the performance of 100 photodetector devices in terms of the responsivity and rise time are displayed in Figure 5e. The statistical distributions reveal a high device yield with a responsivity of $0.255\pm0.01$ A/W and a rise time of $98\pm10$ μs (523 nm, of 23.39 mW/cm$^2$). The fluctuations in the measured values could be attributed to the slight nonuniformity in the thickness of the film across the whole wafer and/or to the nature of the metal-semiconductor contact between the individual devices.

## 4    Conclusion

In conclusion, wafer-scale $Sb_2Te_3$ thin films were synthesized using the ALD process. A detailed structural characterization of the as-grown layers was conducted using a combination of approaches involving HRTEM, Raman, and XPS spectroscopy. The studies indicate good crystal quality, a low density of extended defects, and the expected stoichiometry of the films. Photodetectors were fabricated using standard lithography tools and were shown to possess superior performance, with a broadband photodetection range of from 405 to 1550 nm, an excellent responsivity of 4.287 A/W, and a fast response speed of 98 μs ($t_{rise}$) at 405 nm in the self-powered mode. Such good performance could be attributed to the excellent crystal quality of the ALD-grown $Sb_2Te_3$ films and the presence of the strong built-in electric field at the heterointerface of the two materials. The fabricated devices were found to be quite stable, showing no degradation in performance even after 12 weeks of exposure to the ambient atmosphere. The photodetector shows stable ON/OFF behaviour even at a temperature as high as 473 K. The photocurrent was found to decrease progressively with increasing temperature. This behaviour could be attributed to the concomitant increase in material resistivity and decrease in carrier concentration. The results reported in this study show a viable approach to preparing monolithically integrated materials with unusual band structures, such as $Sb_2Te_3$ topological insulators, with Si, using the low-temperature ALD growth process, resulting in excellent film quality and high-performance and scalable optoelectronic devices using CMOS-compatible processes, paving the way to exciting new opportunities in areas such as integrated photonics, compact imaging and sensing platforms, and on-chip optical communication.

## Supporting Information

Device fabrication procedure, growth mechanism and optical images of $Sb_2Te_3$ thin film, photocurrent switching behaviours, detectivity of devices, and differential charge density diagram.

**Conflict of Interest**

The authors declare no conflicts of interest.

**Acknowledgements**

This work was supported by the Program of Collaborative Research Centers in Germany (Grant No.: SFB 1415). Special thanks to Dr. Heiko Reith, Dennis Hofmann, and Steve Wohlrab for technical assistance and Ronald Uhlemann for the preparation of the schematic illustrations.

# Figures and Tables

Table 1. Comparison of the Key Parameters of the Fabricated Photodetectors with the Literature

| Device materials | Condition | Peak responsivity (A/W) | Peak detectivity (Jones) | Response time ($t_{rise}/t_{fall}$) | Ref. |
|---|---|---|---|---|---|
| Ge | 1550 nm @0 V | 92 | $3.57 \times 10^{11}$ | 22/50 µs | 57 |
| Ge | 1550 nm @0 V | 51.8 | $1.38 \times 10^{10}$ | 23/108 µs | 58 |
| Ge/Si | 1550 nm @-1 V | 0.8 | - | - | 59 |
| Si | 520 nm @4 V | $8.5 \times 10^{-3}$ | - | - | 60 |
| Si | 850 nm @10 V | $7.5 \times 10^{-3}$ | - | - | 61 |
| GaAs | 532 nm@8 V | 46.3 | $2.30 \times 10^{13}$ | 210/374 ms | 62 |
| GaAs | 532 nm@0 V | 0.301 | $2.10 \times 10^{9}$ | 42/140 µs | 63 |
| $Sb_2Te_3$/Si | 405-1550 nm @0 V | 4.3 | $1.65 \times 10^{14}$ | 98/133 µs | This work |

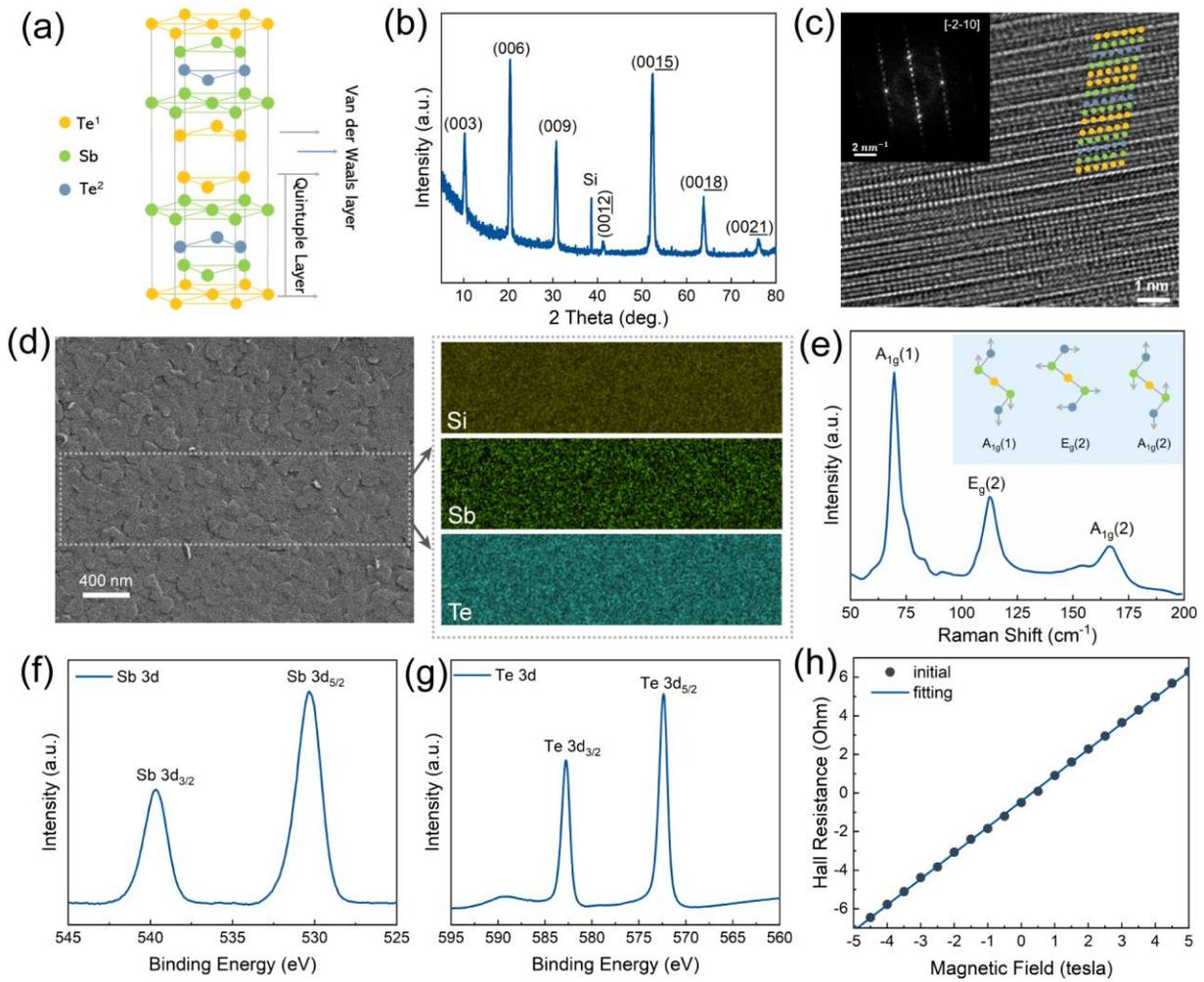

Figure 1. Characterization of the $Sb_2Te_3$ thin films. (a) Atomic structure and (b) XRD pattern of $Sb_2Te_3$. (c) HRTEM image of $Sb_2Te_3$. Inset: Fourier transform kinematical electron diffraction pattern along the [-2-10] direction. (d) Top-view SEM and EDX results of $Sb_2Te_3$. (e) Raman spectra of the $Sb_2Te_3$ thin film. The XPS spectra of (f) Sb 3d and (g) Te 3d, and (h) Hall resistance versus magnetic field in $Sb_2Te_3$ thin film at 300 K.

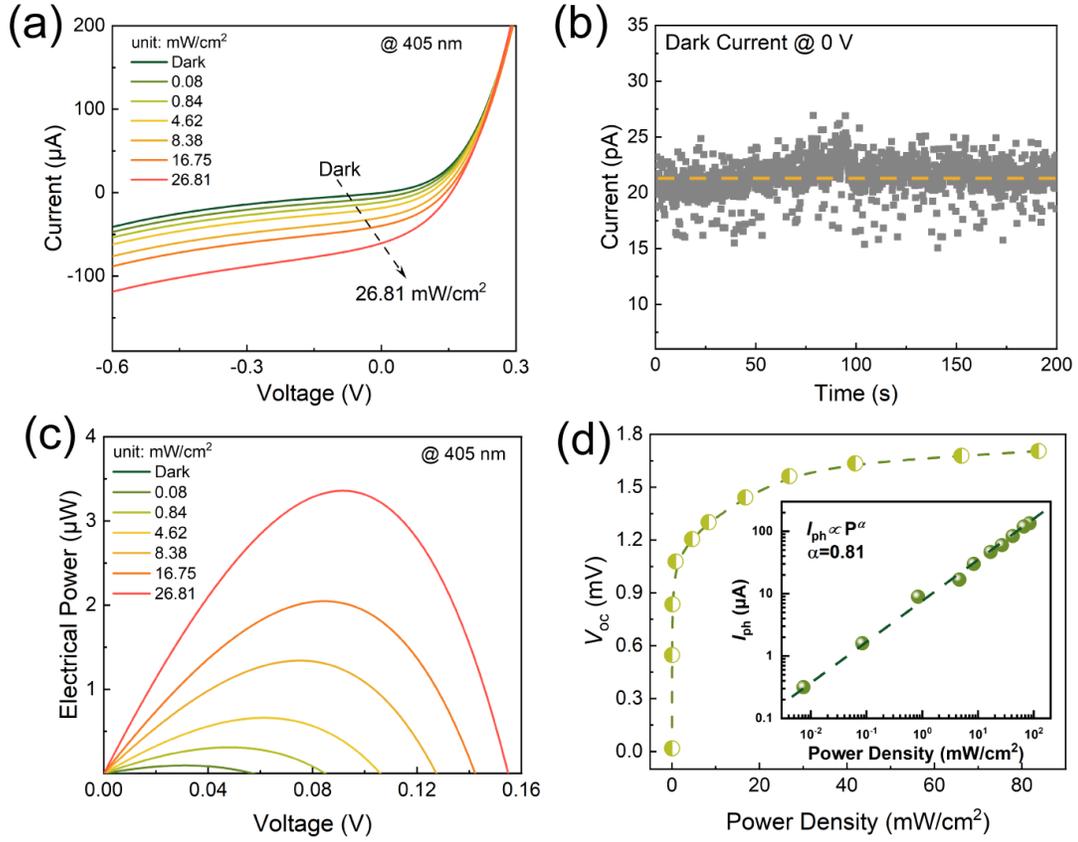

Figure 2. Optoelectronic characteristics of the Sb$_2$Te$_3$/Si heterostructure. (a) $I$-$V$ measurement under different laser power intensities at a wavelength of 405 nm at room temperature. (b) The dark current of the Sb$_2$Te$_3$/Si photodetector device was recorded at 0 V for 200 s. (c) The electrical power extracted from the $I$-$V$ curve. (d) Dependence of the open circuit voltage $V_{oc}$ on incident light power at 405 nm. Inset: photocurrent $I_{ph}$ vs. laser power densities.

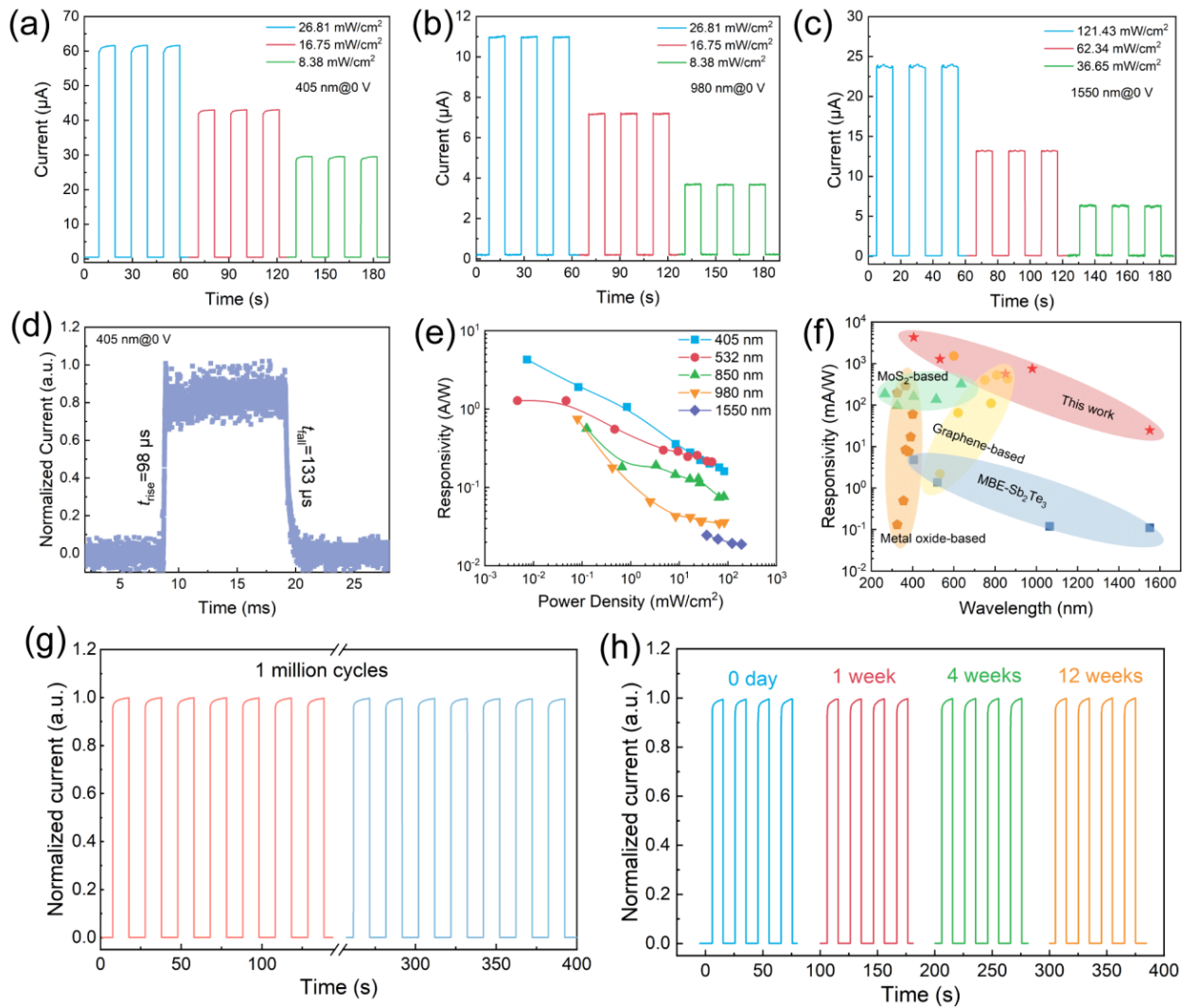

Figure 3. Self-powered photoresponse characteristics of the Sb$_2$Te$_3$/Si heterostructure. Photocurrent switching performance of the photodetector device under (a) 405 nm, (b) 980 nm, and (c) 1550 nm laser illumination with different light power intensities. (d) Photoswitching speed of the Sb$_2$Te$_3$/Si heterostructure under 405 nm. (e) Responsivity of the Sb$_2$Te$_3$/Si photodetector at different wavelengths under 0 bias voltage. (f) Comparison of photoresponsivity of the self-powered photodetector. Operational stability of the photodetector for (g) 1 million cycles test and (h) 12 weeks measurement.

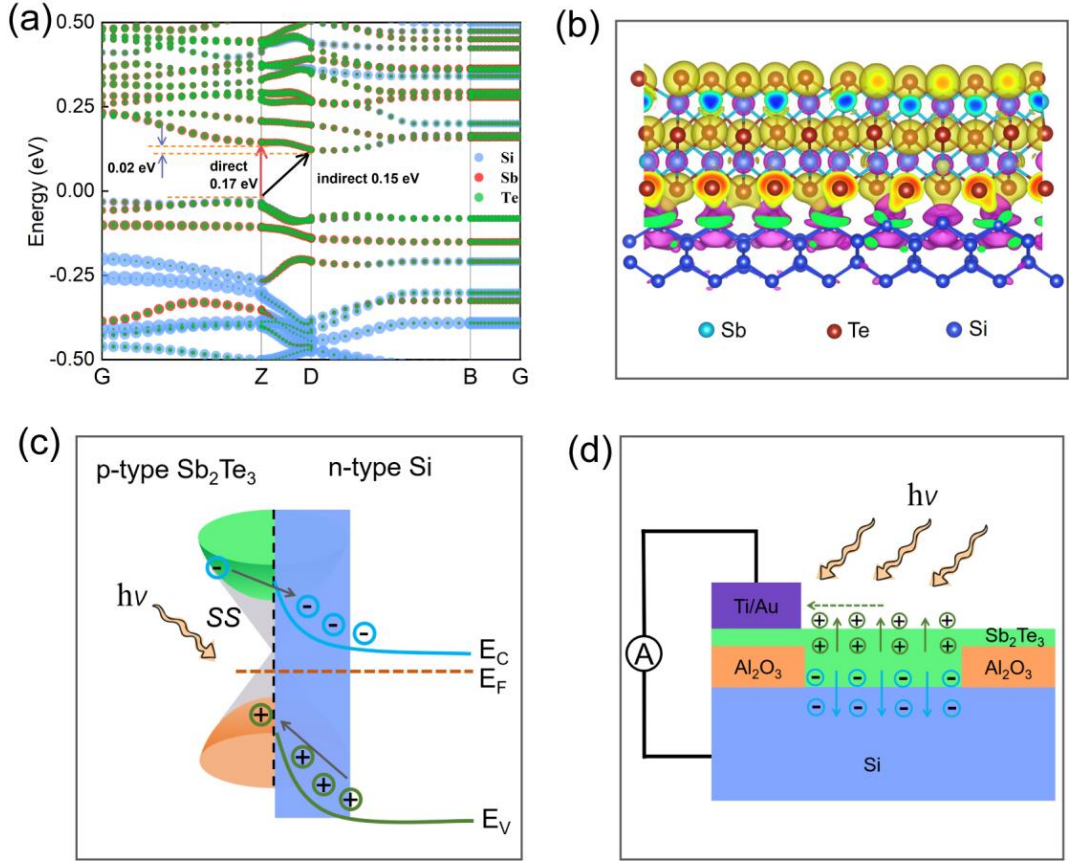

Figure 4. DFT calculation and photocurrent mechanism of the Sb$_2$Te$_3$/Si heterostructure. (a) Energy band structure of the Sb$_2$Te$_3$/Si heterostructure, showing the relative contributions of each element. The equilibrium Fermi level is located at 0 eV. The fundamental indirect transition at the heterointerface is denoted by the black arrow, while the direct transition is denoted by the red arrow. (b) Differential charge density diagram calculated by the DTF simulation. The yellow and purple clouds indicate charge accumulation (+) and depletion (-), respectively. The green part is the cross-section of electronic charges at the edge of a unit cell. (c) The working mechanism of the Sb$_2$Te$_3$/Si photodetector at 0 bias voltage under illumination. *SS* is the surface state of the Sb$_2$Te$_3$ topological insulator thin film. (d) Schematic illustration of the transportation and generation of carriers across the Sb$_2$Te$_3$/Si heterojunction device.

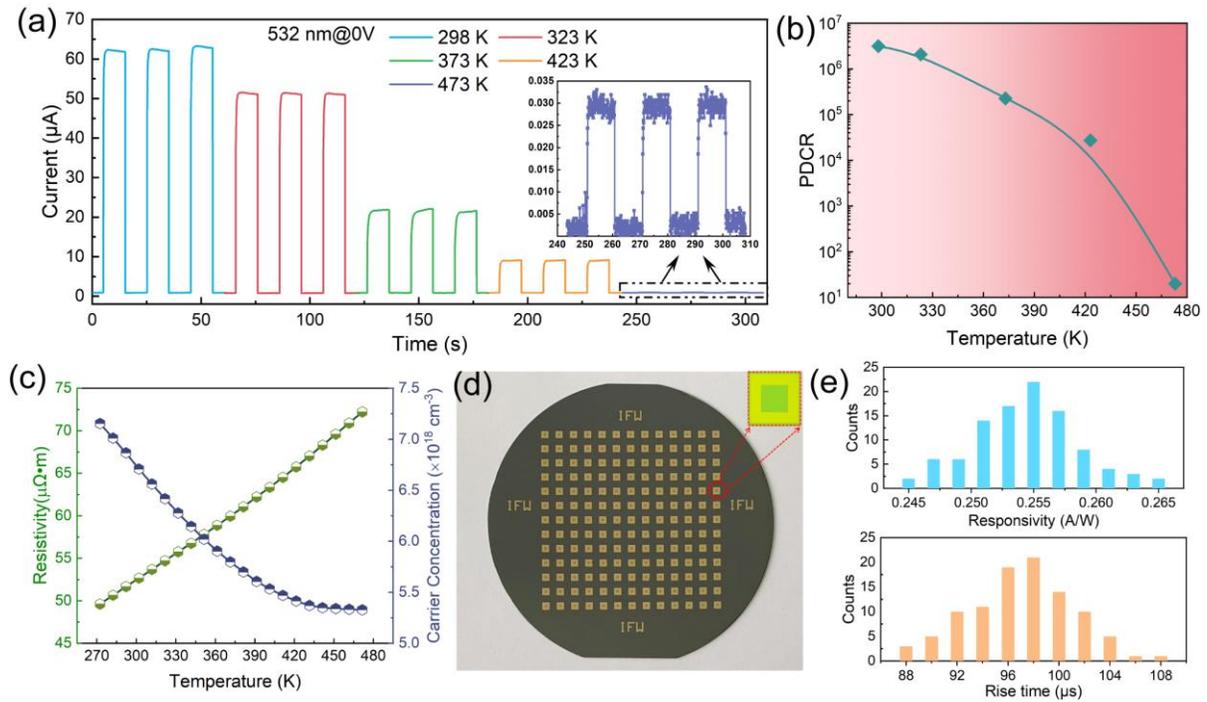

Figure 5. Temperature dependence performance of the Sb$_2$Te$_3$/Si heterostructure photodetector. (a) Photoswitching behaviours (the inset shows the photoswitching behaviours at 473 K) and (b) Photo-to-dark current ratio (PDCR) of the photodetector under a laser of 523 nm with an illumination power density of 23.39 mW/cm$^2$ at temperatures ranging from 298 to 473 K. (c) The transport property, including resistivity and carrier concentration of the ALD Sb$_2$Te$_3$ thin film at the temperatures ranging from 273 to 473 K. (d) Optical image of patterned Sb$_2$Te$_3$ on a 4-inch n-type Si wafer substrate. (e) Statistical results of the responsivity and rise time of detectors under a laser of 523 nm with an illumination power density of 23.39 mW/cm$^2$.

# Supporting Information

# Wafer-Scale Growth of Sb₂Te₃ Films via Low-Temperature ALD for Self-Powered Photodetector

*Jun Yang[1,2*], Jianzhu Li[3], Amin Bahrami[1], Noushin Nasiri[4], Sebastian Lehmann[1], Magdalena Ola Cichocka[1], Samik Mukherjee[1*], Kornelius Nielsch[1,2*]*

[1] Institute for Metallic Materials, Leibniz Institute of Solid State and Materials Science, 01069 Dresden, Germany

[2] Institute of Materials Science, Technische Universität Dresden, 01062 Dresden, Germany

[3] School of Materials Science and Engineering, Harbin Institute of Technology (Weihai), Weihai, West Road 2, Weihai, Shandong 264209, China

[4] School of Engineering, Faculty of Science and Engineering, Macquarie University, Sydney, NSW 2109, Australia

Corresponding authors:

Jun Yang (j.yang@ifw-dresden.de)

Samik Mukherjee (s.mukherjee@ifw-dresden.de)

Kornelius Nielsch (k.nielsch@ifw-dresden.de)

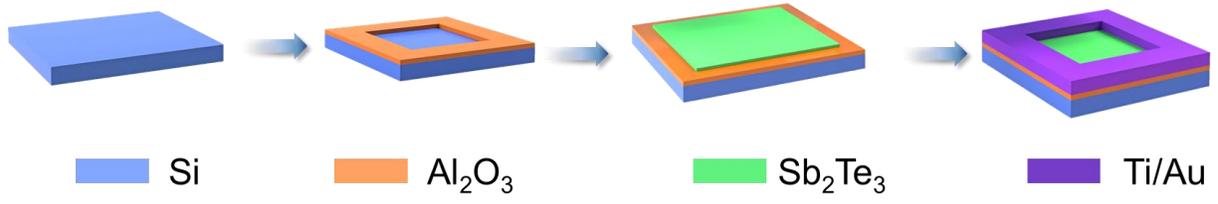

Figure S1. The fabrication process of Sb₂Te₃/Si heterostructure photodetector.

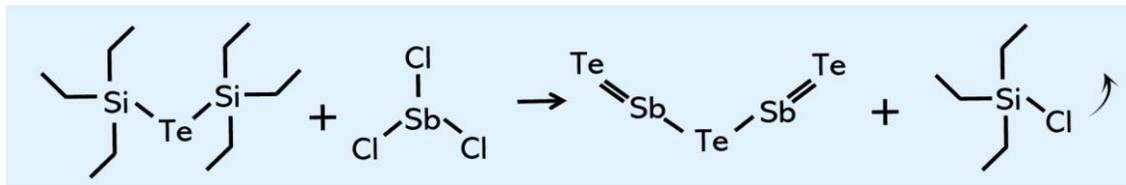

Figure S2. Schematic diagram of the reaction mechanism for ALD-Sb₂Te₃ using (Et₃Si)₂Te and SbCl₃

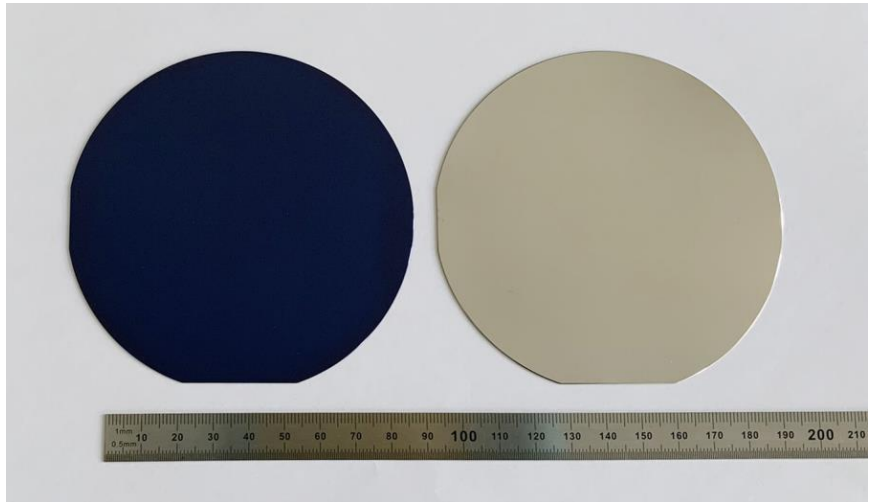

Figure S3. The optical images of 4-inch scale Sb₂Te₃ thin film via ALD (Left: bare SiO₂/Si. Right: after deposition).

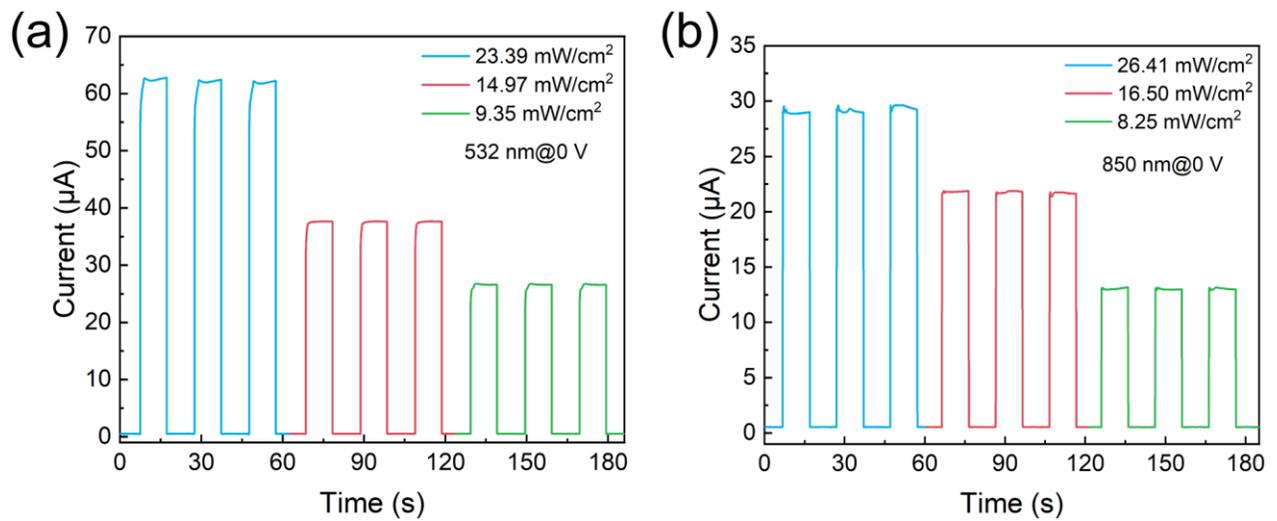

Figure S4. Photocurrent switching performance of photodetector device under (a) 532 nm and (b) 850 laser illumination with different light power intensities, respectively.

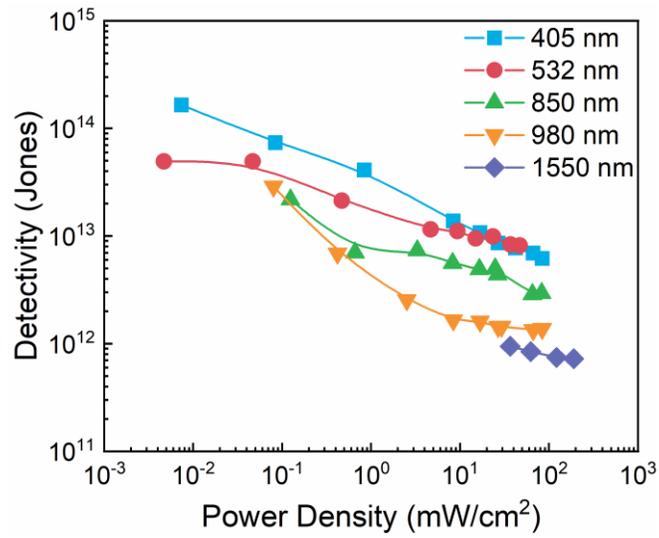

Figure S5. Detectivity of Sb₂Te₃/Si photodetector at different wavelengths under 0 bias voltage.

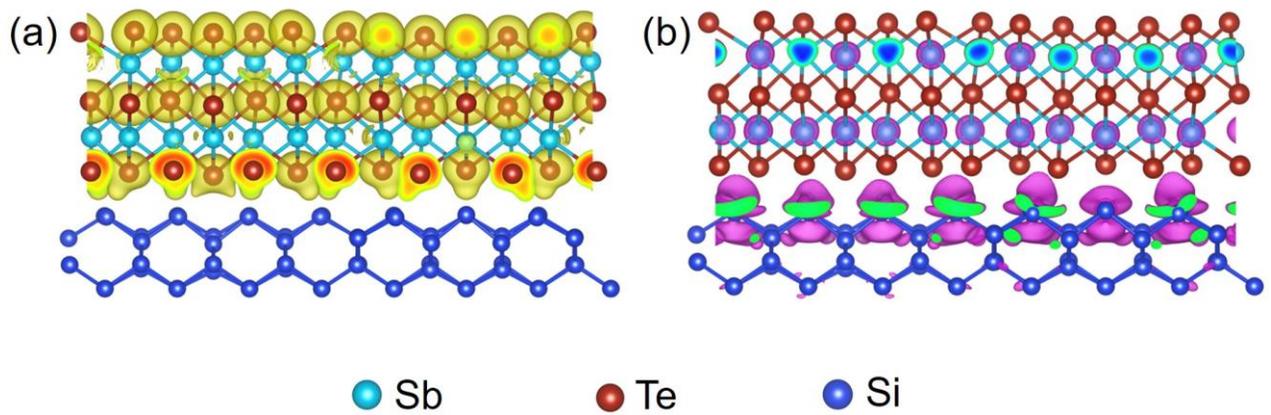

Figure S6. Differential charge density diagram of Sb₂Te₃/Si heterostructure: (a) only charge accumulation (+) and (b) only charge depletion (-).

**Discussions on the interaction of electromagnetic waves with electron/hole systems:**

A popular method to elucidate the interaction of electromagnetic waves (EMW) with electron/hole system is by using the semi-classical model, wherein the radiation is treated classically and the energy levels are obtained using the time-dependent Schrödinger equation: $\hat{H}\Psi(r,t) = i\hbar.\partial\Psi(r,t)/\partial t$. The Hamiltonian $\hat{H}$ is a combination of the Laplacian operator $\nabla^2$ and the potential energy operator $\hat{V}$. The equation is often solved using the separation of variable technique: $\Psi(r,t) = \psi(r).\phi(t)$, on the assumption that $\hat{V}$ depends only on the spatial coordinates (r) and not on time (t). The time-independent part of the Schrödinger equation gives the energy eigenvalues of the electrons in the state $n$ ($E = E_n$) and the eigenstate $\psi(r)$ wherefrom all physical properties can be deduced. The time-dependent part is a first-order differential equation $i\hbar.\partial\phi(t)/\partial t = E\phi(t)$, with solutions given by $\phi(t) = \exp(-iE_nt/\hbar)$. Since the time-dependent part of the Schrödinger equation always bears the same form, the total electron wavefunction can be written as: $\Psi_n = \psi_n\exp(-iE_nt/\hbar)$.

More complex systems, like electrons in a crystal, usually possess several stationary states described by the total wavefunction $\Psi_0$, $\Psi_1$, $\Psi_2$,..... with discrete energies $E_0$, $E_1$, $E_2$,...... For a transition between two non-degenerate $m$ and $n$ states caused by a photon of adequate frequency, the $\Psi_m$ and $\Psi_n$ wavefunctions describing those states are a solution of the Schrödinger equation and so is their linear combination: $\Psi = c_m(t)\Psi_m + c_n(t)\Psi_n$. The initial state $\Psi_m$ is described by $c_m = 1/ c_n = 0$ and the final state $\Psi_n$ is described by $c_m = 0/ c_n = 1$.During photon absorption the coefficients change from $c_m = 1 \rightarrow 0$ to $c_n = 0 \rightarrow 1$, while the opposite happens during photon emission.[1]

To understand how the EMW-induced transition of an electron between the states $\Psi_m$ and $\Psi_n$ take place in a crystal or a solid-state optoelectronic device, we need to consider that the time-dependent radiation field associated with the EMW is a perturbation $\hat{H}'(r,t)$ that needs to be added to the Hamiltonian of the unperturbed system $\hat{H}_o$ to get $\hat{H} = \hat{H}_o + \hat{H}'$.If the perturbating potential of the EMW is small compared to the atomic potential within the crystal/device then $(\hat{H}_o + \hat{H}')\Psi = i\hbar.\partial\Psi/\partial t$ , which can be solved using the perturbation theory. It can be shown that the rate of transition from the initial state $m$ to the final state $n$ is:[1,2]

$$\frac{dc_n}{dt} = \frac{c_m}{i\hbar} \int \Psi_n^* \hat{H}'\Psi_m \, d\upsilon$$

wherein explicit forms of $\Psi_m$, $\Psi_n$ wavefunctions and the $\hat{H}'$ Hamiltonian are needed to evaluate the intensity of a radiative transition. To elucidate further the interaction between the electric/magnetic components of the EMW and the electric/magnetic dipole of the system must be considered. For sake of brevity, let's consider an electric dipole moving along the '$x$' direction, such that the dipole moment is $\mu_x = ex$, where e is the electronic charge and $x$ is the displacement from the equilibrium position. $\hat{H}' =$

$-\mu_x E_x = -ex E_x$, where $E_x$ is the amplitude of the electric field of a polarized EMW traveling along the x-direction and is given by $E_x = E_x^o\left(e^{i2\pi\nu t} + e^{-i2\pi\nu t}\right)$, where $E_x^o$ is the maximum field amplitude and $\nu$ is the frequency. Therefore,

$$c_n = E_x^o \int \Psi_n^* \, \mu_x \Psi_m \, d\nu \left[\frac{1 - e^{i(E_n - E_m + h\nu)t/\hbar}}{E_n - E_m + h\nu} + \frac{1 - e^{i(E_n - E_m - h\nu)t/\hbar}}{E_n - E_m - h\nu}\right]$$

If the energy of the initial state $m$ is lower than that of the final state $n$, then $E_n - E_m$ is positive for the $\Psi_m \to \Psi_n$ the transition caused by photon absorption and will be negative for a $\Psi_m \leftarrow \Psi_n$ the transition caused by the photon emission. For the photon absorption process which takes place in the photodetector discussed in this work, $E_n - E_m \cong h\nu$, the second denominator within the square bracket in the expression for $c_n$ becomes vanishingly small such that the entire first term can be neglected. Under these conditions and integrating over all possible range of electromagnetic frequencies, the probability of finding the system in the upper state $n$ as a result of photon absorption is given by

$$|c_n|^2 = c_n^* \, c_n = 4(E_x^o)^2 (\mu_x^{nm})^2 \int\limits_{-\infty}^{+\infty} \frac{\sin^2[(E_n - E_m - h\nu) \, \pi t/\hbar]}{(E_n - E_m - h\nu)^2} d\nu = \frac{(E_x^o)^2}{\hbar^2} (\mu_x^{nm})^2 t$$

Hence, the probability of a system in the initial state $\Psi_m$ making a transition to the final state $\Psi_n$ owing to photon absorption is proportional to the transition dipole moment, the square of the incident electric field amplitude, and the irradiance time. To compare with experimental results, it is more convenient to express the field strength by the energy density ($\rho$) that is irradiated by the EMW: $\rho = (E_x^o)^2/2\pi$. For an isotropic radiation field where the $x,y,z$ components of the radiation-diploe interaction are equal, the transition probability per unit time to the final state $n$ as a result of photon absorption is given by

$$\frac{d[c_n^* \, c_n]}{dt} = \frac{8\pi^3}{3h^2} \mu_{nm}^2 \rho(\nu) = B_{nm}\rho(\nu), \text{ where } \mu_{nm}^2 = \sum_{i=x,y,z} (\mu_i^{nm})^2$$

$B_{nm} = (8\pi^3/3h^2)\mu_{nm}^2$ is the Einstein coefficient of induced absorption. On changing the initial condition with a system in an upper state with $c_m = 0$ and $c_n = 1$, the analysis for induced absorption would be identical to the emission of radiation from an initial excited state $\Psi_n$ to the final state $\Psi_m$ induced by EMW. The coefficient of induced emission $B_{mn}$ from an excited state is identical to that of $B_{nm}$.

# TOC

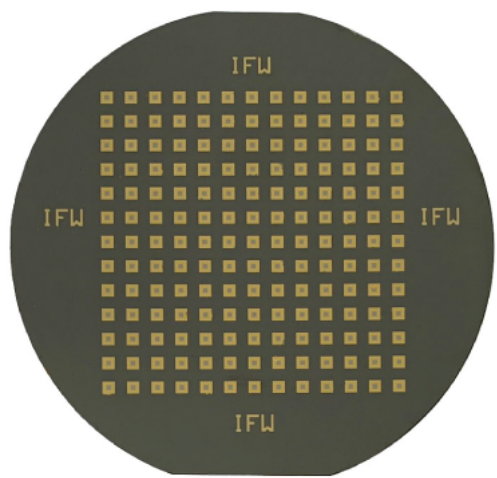
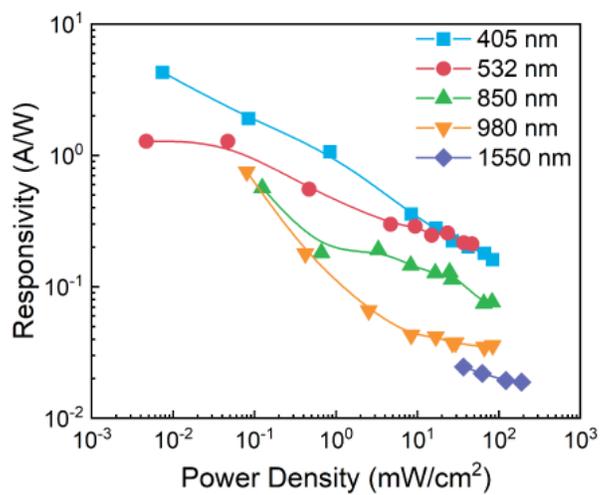